\documentclass[traditabstract]{aa}  
\usepackage{graphicx}
\usepackage{natbib}
\usepackage{txfonts}
\usepackage{color}
\usepackage{ulem}
\usepackage{textcomp}
\begin{document}

   \title{Photospheric activity of the Sun with VIRGO and GOLF}
   \subtitle{Comparison with standard activity proxies}
   
   \author{D. Salabert  \inst{1,2}
          \and
           R.~A. Garc\'ia \inst{1,2}
	 \and
	  A. Jim\'enez\inst{3,4} 
	  \and 
	   L. Bertello\inst{5}
	  \and
	  E. Corsaro\inst{1,2,3,4,6}
	  \and
	   P.~L. Pall\'e\inst{3,4}
	           }

         \institute{IRFU, CEA, Universit\'e Paris-Saclay, F-91191 Gif-sur-Yvette, France\\
            \email{david.salabert@cea.fr}
            \and
        Universit\'e Paris Diderot, AIM, Sorbonne Paris Cit\'e, CEA, CNRS, F-91191 Gif-sur-Yvette, France
          \and
    	Instituto de Astrof\'isica de Canarias,  E-38200 La Laguna, Tenerife, Spain
         \and
      	Departamento de Astrof\'isica, Universidad de La Laguna, E-38205 La Laguna, Tenerife, Spain
	\and
          National Solar Observatory, 3665 Discovery Drive, Boulder, CO 80303, USA
          \and
          INAF - Observatorio Astrofisico di Catania, Via S. Sofia 78, I-95123 Catania, Italy} 

   \date{Received XXXXX; accepted XXXXX}
 
  \abstract
  {We study the variability of solar activity using new photospheric proxies originally developed for the analysis of stellar magnetism with the CoRoT and {\it Kepler} photometric observations. These proxies are obtained by tracking the temporal modulations in the observations associated to the spots and magnetic features as the Sun rotates. We analyze here 21 years, spanning solar cycles 23 and 24, of the observations collected by the space-based photometric VIRGO and radial velocity GOLF instruments on board the SoHO satellite. The photospheric activity proxy $S_\text{ph}$ is then calculated for each of the three VIRGO photometers as well as the associated $S_\text{vel}$ proxy from the radial velocity GOLF observations. Comparisons with several standard solar activity proxies sensitive to different layers of the Sun demonstrate that these new activity proxies, $S_\text{ph}$ and $S_\text{vel}$, provide a new manner to monitor solar activity. We show that both the long- and short-term magnetic variabilities respectively associated to the 11-year cycle and the quasi-biennial oscillation are well monitored, and that the magnetic field  interaction between the subsurface, the photosphere, and the chromosphere of the Sun, was modified between Cycle~24 and Cycle~23. Furthermore, the photometric proxies show a wavelength dependence of the response function of the solar photosphere among the three channels of the VIRGO photometers, providing inputs for the study of the stellar magnetism of Sun-like stars.
}  
  

   \maketitle
%
%________________________________________________________________

%________________________________________________________________
\section{Introduction}
The space-based, Sun-as-a-star Variability of Solar Irradiance and Gravity Oscillations  \citep[VIRGO;][]{frohlich95} and Global Oscillations at Low Frequency \citep[GOLF;][]{gabriel95} instruments on board the Solar and Heliospheric Observatory \citep[SoHO;][]{domingo95} satellite were designed to study the internal structure and rotation of the Sun by measuring the low-degree oscillations over the entire disc in intensity and radial velocity, respectively \citep[e.g., in the following early references][]{turck97,frohlich97}. The first reliable signatures of the existence of gravity modes in the Sun (and by consequence in Sun-like stars) were also detected with the GOLF instrument \citep{garcia07,fossat17}. Moreover, by adding three polarizing elements at the entrance of the instrument, GOLF was able to measure the two circular polarized components of the solar light and thus to estimate the solar mean magnetic field (SMMF). Unfortunately, due to a malfunction of the motors moving the polarizers, only 26 days of SMMF observations were acquired. From those observations, an average value of 0.120\,$\pm$\,0.002 G was determined \citep{garcia99}.

Long and continuous helioseismic observations offer a unique opportunity to monitor and study solar activity at different temporal scales. The response of the acoustic oscillations to solar activity, through the variability of the eigenfrequencies, was proven to provide insights on the structural and magnetic changes in the sub-surface layers of the Sun along the solar cycle \citep[see][and references therein]{basu12,salabert15}. However, while the frequency shifts varied with very high levels of correlation with surface activity over the last three solar cycles 21, 22, and 23 \citep{chaplin07}, significant differences were observed between frequency shifts and solar activity during the unusual long and deep minimum of Cycle 23 \citep{salabert09,tripathy10} related to magnetically weaker shallow layers of the Sun. In addition, a faster temporal modulation of the acoustic frequency shifts with a period of about two years was also observed \citep{fletcher10}. This quasi-biennial oscillation (QBO) is measured in various activity proxies \citep[see e.g.,][and references therein]{bazi15}, and is characterized by intermittence in periodicity and
with amplitudes that vary with time, being largest at solar maximum.
In addition, the other acoustic parameters, like the mode powers and lifetimes, were also observed to vary with magnetic activity in relation with the 11-year solar cycle \citep{chaplin00,salabert03}.

% -------------------
\begin{figure*}[tbp]
\begin{center} 
\includegraphics[width=\textwidth,angle=0]{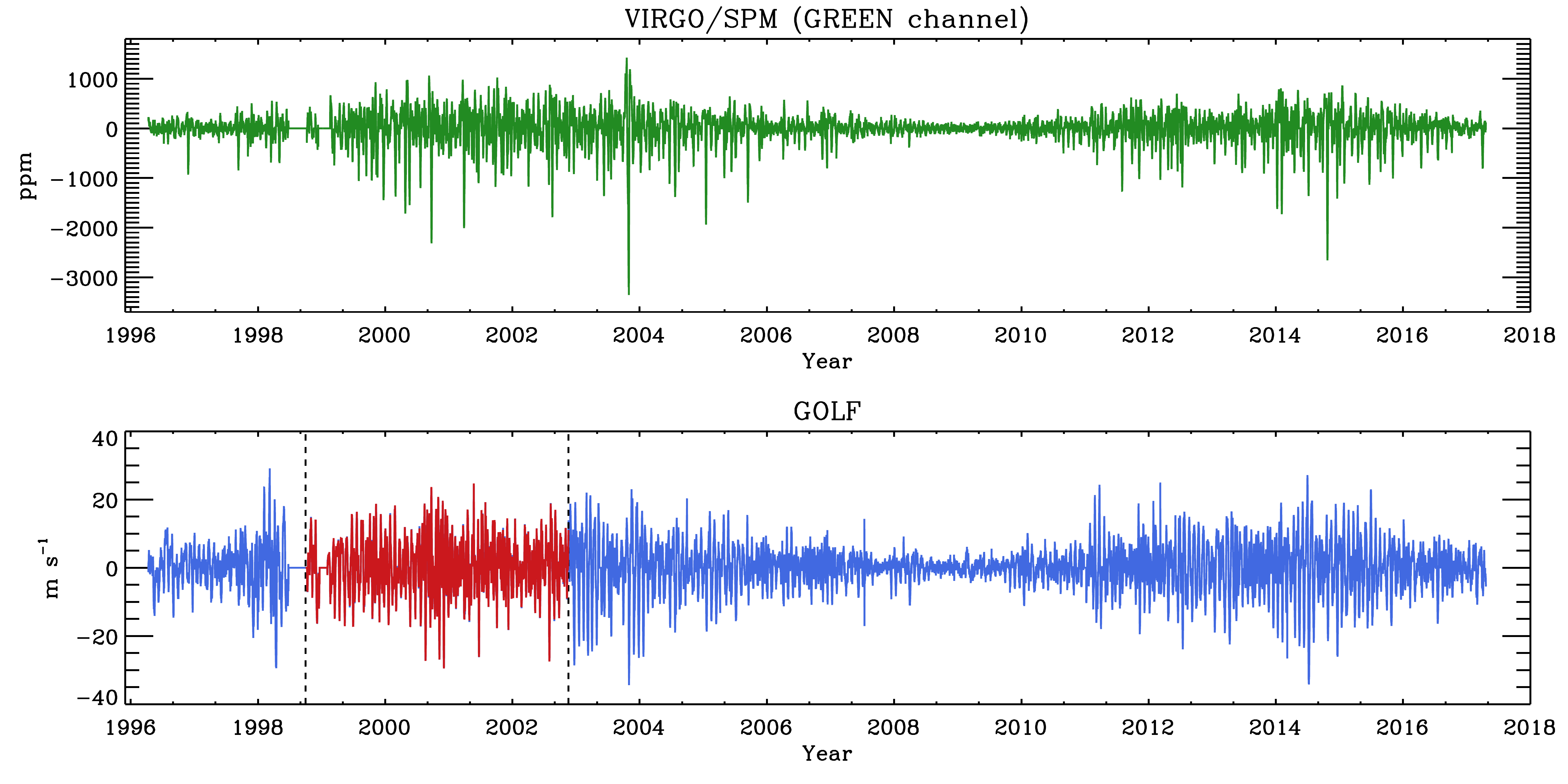}
\end{center}
\caption{\label{fig:fig1} 
21 years of space-based observations of the Sun collected by the photometric VIRGO/SPM (top panel, here the \textsc{green} channel) and the radial velocity GOLF (bottom panel) instruments on board the SoHO satellite between 1996 April 11 and 2017 April 11. The missing data shown at zero between the years 1998 and 1999 correspond to the two temporary losses of SoHO. The GOLF red-wing period from mid 1998 to end 2002, denoted by the two vertical dashed lines, is represented in red (see Section~\ref{sec:obs}).}
\end{figure*} 
% -------------------

Space-based photometric and radial velocity helioseismic observations provide an unexplored way yet to monitor solar activity. In the stellar context, \citet{garcia10} showed, in the case of the F-star HD\,49933 observed with the Convection, Rotation, and planetary Transits \citep[CoRoT;][]{baglin06} space telescope, that the fluctuations associated to the presence of spots or magnetic features rotating on the surface of the star can be analyzed to derive a global proxy of stellar magnetic activity. Nevertheless, as the variability in the data can have different origins with various timescales, such as convective motions, oscillations, stellar companion, or instrumental problems, the rotation period of the star needs to be taken into account in calculating such a magnetic activity proxy. \citet{mathur14a} demonstrated that the measured fluctuations estimated as the standard deviations calculated over sub series of length $5\times P_\text{rot}$, where $P_\text{rot}$ is the rotation period of the star in days, provide a global proxy only related to magnetism. It is referred since then to the so-called photospheric activity proxy $S_\text{ph}$. Initially developed to be applied to the photometric observations collected by the {\it Kepler} \citep{borucki10} and CoRoT satellites to study the signature of stellar magnetic activity of main-sequence stars \citep{garcia14,mathur14b,ferreira15,salabert17}, the $S_\text{ph}$ proxy was used to show that the activity of seismic solar analogs is comparable to the Sun, within the maximum-to-minimum temporal variations of the 11-year solar cycle \citep{salabert16}. Moreover, \citet{salabert16} show the complementarity between the solar photospheric $S_\text{ph}$ proxy and the chromospheric activity such as the Ca K-line emission index. Furthemore, the photospheric $S_\text{ph}$ proxy can be easily estimated for a large number of stars with known rotation period from space photometric observations, unlike the standard chromospheric $\mathcal{S}$ index \citep{wilson78} whose estimation requires a lot of time of ground-based telescopes to collect enough spectroscopic data for each individual bright target only. We note that the $S_\text{ph}$ index is dependent on the inclination angle of the rotation axis in respect to the line of sight and thus provides a lower limit of stellar activity, assuming that the starspots are formed over comparable ranges of latitude as in the Sun. Nevertheless, if the distribution of the spin orientation in space is random, then the most observed inclination angle would be close to 90$^{\circ}$, hence perpendicular to the line of sight \citep[see][]{corsaro17}. This implies, assuming this hypothesis is verified, that the $S_\text{ph}$ index thus measured represents then the actual level of activity for most of the stars and not a lower limit.

In this work, we analyzed 21 years of the space photometric VIRGO and radial velocity GOLF observations to show that such photospheric proxies can be used to monitor the solar activity and its long- and short-term temporal evolutions. In Section~\ref{sec:obs}, we describe the set of observations used in this analysis. In Section~\ref{sec:activity}, we analyze the temporal variations of these photospheric activity proxies over the solar cycles 23 and 24 and we compare them to standard activity proxies. In Section~\ref{sec:wave}, we study the wavelength dependence of the photospheric $S_\text{ph}$ proxy derived from the VIRGO observations. Conclusions are presented in Section~\ref{sec:conclusion}.
%________________________________________________________________

 %________________________________________________________________
\section{Observations}
\label{sec:obs}
The space-based VIRGO and GOLF instruments on board SoHO are collecting continuous observations of the Sun since the beginning of 1996, with temporal cadences of 60 and 10~seconds respectively. However, we note that two temporary losses of the SoHO spacecraft result in two extended gaps in the VIRGO and GOLF acquisition. The first gap of about 100 days happened during the summer of 1998 after a bad maneuver of the SoHO rotation. The second gap, about a month long, occurred in January 1999 while a new software was being uploaded to the spacecraft that caused the SoHO gyroscopes to fail. In this work, we used a total of 21 years of VIRGO and GOLF data from 1996 April 11 to 2017 April 11 spanning solar cycles 23 and 24. The associated duty cycles are 94.7\% and 96.6\% respectively.

 The VIRGO instrument is composed of three Sun photometers (SPM) at 402~nm (\textsc{blue} channel), 500~nm (\textsc{green} channel), and 862~nm (\textsc{red} channel). The VIRGO/SPM photometric observations were calibrated as described in \citet{jimenez02}. In addition, a composite photometric time series was also obtained by combining the observations from the \textsc{green} and \textsc{red} VIRGO channels, whose combination has the closest bandwidth to the {\it Kepler} instrument \citep[see][]{basri10}. This \textsc{composite} series should be used when comparing the Sun with the stars observed by {\it Kepler}.
 
The GOLF instrument was designed to measure the Doppler wavelength shift integrated over the solar surface in the D1 and D2 Fraunhofer sodium lines at 589.6 and 589.0 nm respectively. Unfortunately, due to a malfunction in its polarization mechanism, the GOLF instrument has not been working in its nominal configuration since shortly after the launch of SoHO \citep{garcia04}. Instead, only one side of the sodium doublet is observed, from which a proxy of the Doppler velocity signal is produced \citep{garcia05}.  Moreover, the GOLF observations have been collected from each side of the doublet as follows:  in the blue-wing configuration from 1996 April 11 until 1998 June 24 (i.e., 805~days), and later on from 2002 November 19 until now (the so-called blue-wing periods); in the red-wing configuration of the sodium doublet between 1998 October 30 until 2002 November 18 (i.e., 1481~days) (the so-called red-wing period). 

The VIRGO/SPM and GOLF instruments are sensitive to the conditions at different heights in the solar atmosphere. 
 In the case of the photometric VIRGO/SPM observations, the response functions of the three channels peak around the base of the photosphere \citep{fligge98}, the \textsc{blue} and \textsc{green} channels at $-20$\,km, and the \textsc{red} channel at $+10$\,km \citep{jimenez05}. On the other hand, \citet{chano07} showed that the response function of the Doppler velocity GOLF measurements in the sodium Fraunhofer line is different between the blue- and the red-wing configurations, with averaged heights in the solar atmosphere of 322\,km and 480\,km respectively. Then, the measurements obtained in the red-wing configuration originate higher up in the solar atmosphere with larger contributions from the chromosphere. 

In order to monitor the long-lived features on the solar surface, the VIRGO/SPM and GOLF data were processed and high-pass filtered through the {\it Kepler} Asteroseismic Data Analysis and Calibration Software \citep[KADACS,][]{garcia11}. We recall that a great care is needed in the choice of the high-pass filter as it affects the absolute values of the measured photospheric activity proxies. Here, a cut-off of 70 days was used, the impact of the filter being thus within the errors as shown in \citet{salabert16}. Finally, the VIRGO/SPM and GOLF data were rebinned at 30~min to mimic the  {\it Kepler} long-cadence data having a temporal sampling of 29.4244~min. Figure~\ref{fig:fig1} shows the 21 years of the photometric VIRGO/SPM (\textsc{green} channel) and radial velocity GOLF observations thus obtained and analyzed in this work.

% -------------------
\begin{figure}[tbp]
\begin{center} 
\includegraphics[width=0.5\textwidth,angle=0]{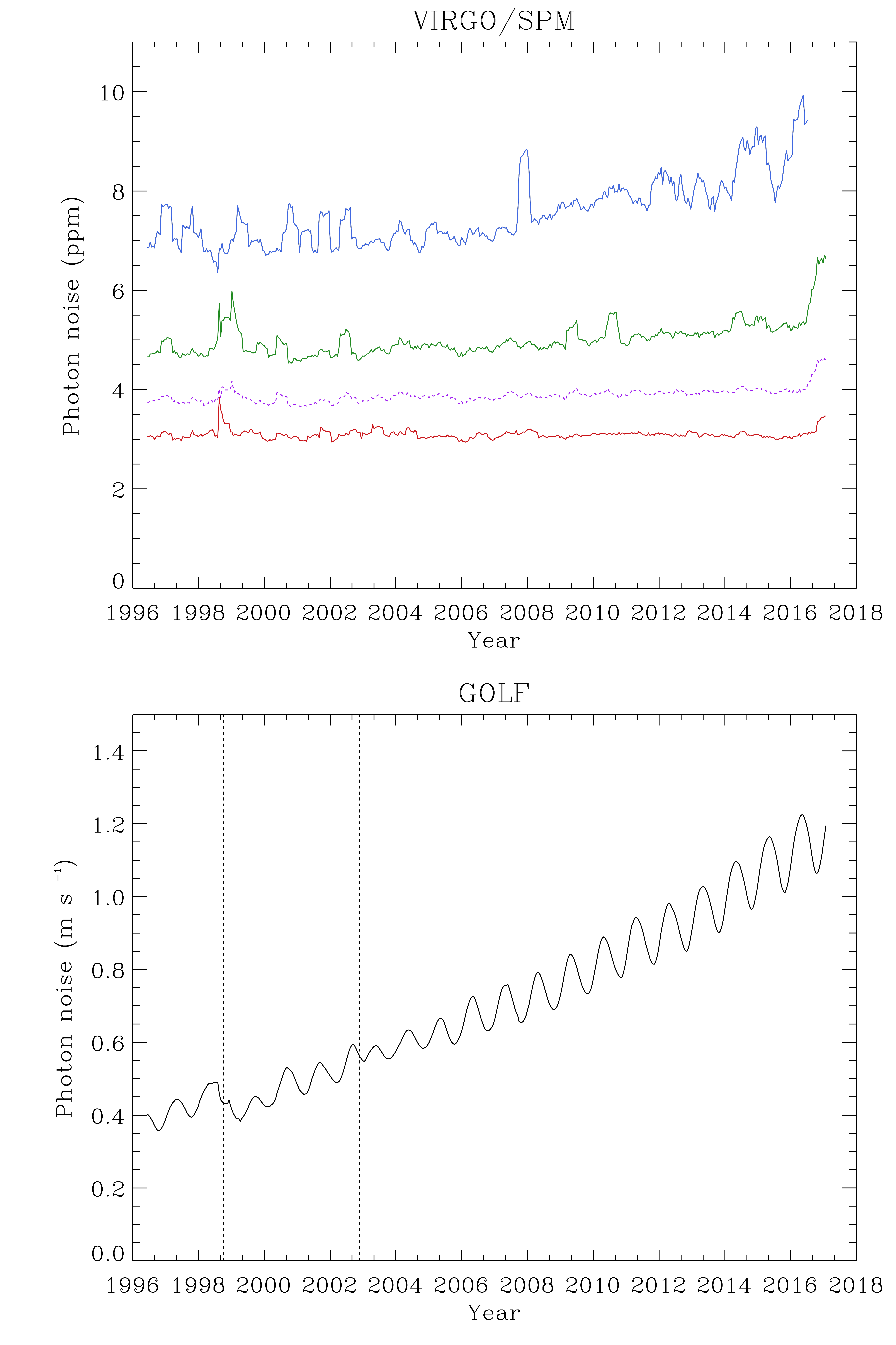}
\end{center}
\caption{\label{fig:fig2} 
{\it Top panel}: Photon noise in ppm as a function of time of the photometric VIRGO/SPM observations (color-coded \textsc{blue}, \textsc{green}, and \textsc{red} channels, and \textsc{composite} in dotted purple). {\it Bottom panel}: Same as the top panel but for the radial velocity GOLF observations in m s$^{-1}$.  The vertical dashed lines indicate the GOLF red-wing period from mid 1998 to end 2002.
}
\end{figure} 
% -------------------

% -------------------
\begin{figure*}[tbp]
\begin{center} 
\includegraphics[width=0.49\textwidth,angle=0]{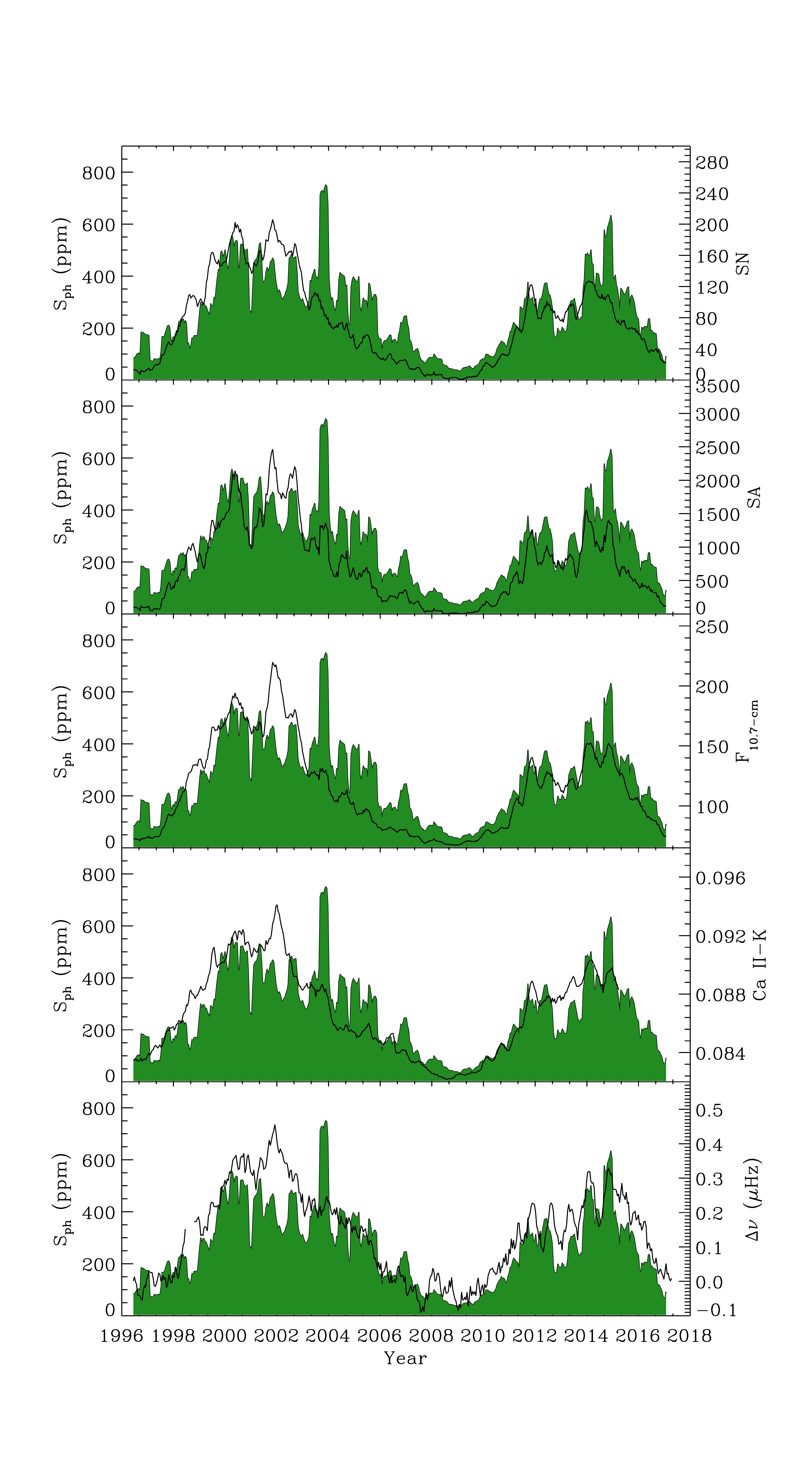}
\includegraphics[width=0.49\textwidth,angle=0]{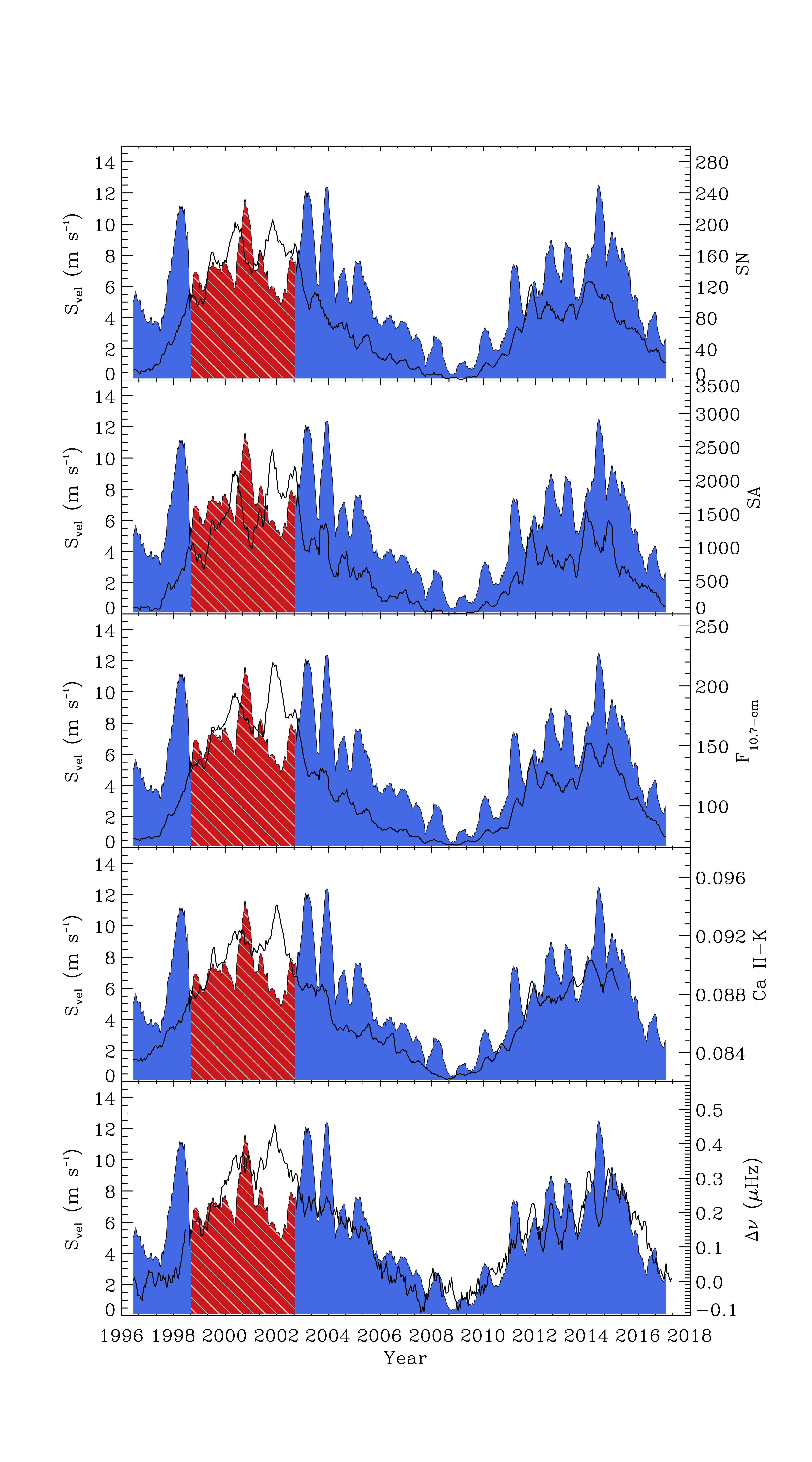}
\end{center}
\caption{\label{fig:fig3} 
Magnetic activity proxies $S_{\text{ph}, \textsc{green}}$ (in ppm) and  $S_\text{vel}$ (in m s$^{-1}$) measured from the photometric VIRGO/SPM \textsc{green} channel (green filled, left panels) and the radial velocity GOLF (blue filled, right panels) observations as a function of time and compared to standard indices of solar activity (solid black lines). From top to bottom: the total sunspot number (SN); the total sunspot area (SA); 
the 10.7-cm radio flux (\text{F$_\text{10.7-cm}$}) in 10$^{-22}$~s$^{-1}$~m$^{-2}$~Hz$^{-1}$;
the Ca \textsc{ii}\,K-line line emission index in $\AA$;  and the mean frequency shifts of the $l=0,1$, and 2 acoustic oscillations ($\Delta\nu$) in $\mu$Hz. The red shaded areas on the right-hand panels represent the period when GOLF was observing in the red wing of the sodium line.}
\end{figure*} 
% -------------------

% -------------------
\begin{figure}[tbp]
\begin{center} 
\includegraphics[width=0.5\textwidth,angle=0]{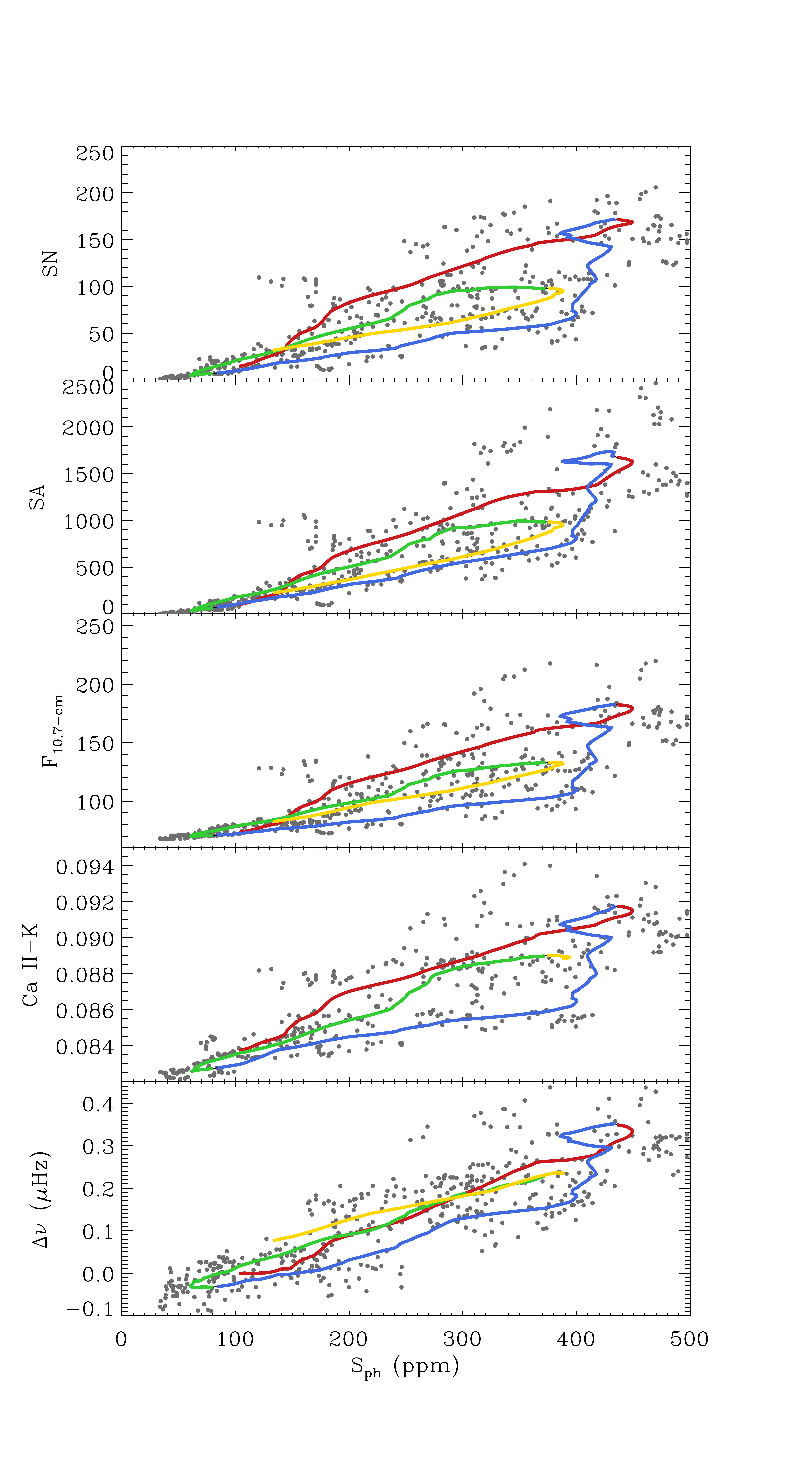}
\end{center}
\caption{\label{fig:fig5} 
Standard proxies of solar activity as a function of the photospheric magnetic proxy of the Sun, $S_{\text{ph},\,\textsc{green}}$ (in ppm), measured from the VIRGO/SPM \textsc{green} channel observations. From top to bottom, the y axis represents the sunspot number, the sunspot area, the 10.7-cm radio flux, the Ca \textsc{ii}\,K emission,  and the mean frequency shifts  of the $l=0,1$, and 2 acoustic oscillations. Individual measurements are shown in gray dots and the solid lines represent a smooth over 2.5 years in order to remove short-term variations for illustrative purpose. The different colors correspond to different phases in solar activity: in red and blue, the rising and declining phases of Cycle~23 respectively; in green and yellow, the rising and declining phases of Cycle~24 respectively.
}
\end{figure} 
% -------------------

%----------------
\begin{table*}[ht]
\begin{minipage}{\textwidth}
\caption{Spearman's correlation coefficients between the photometric VIRGO/SPM $S_{\text{ph}}$  and the radial velocity GOLF $S_{\text{vel}}$ magnetic proxies and standard indices of solar activity over 21 years during Cycles 23 and 24.}
\label{table:table1}      
\centering               
\renewcommand{\footnoterule}{}              
\begin{tabular}{c c c c c c c c c c c}        
\hline\hline   
 Activity proxy &  $S_{\text{ph}, \textsc{blue}}$ & $S_{\text{ph}, \textsc{green}}$ & $S_{\text{ph}, \textsc{red}}$ & $S_{\text{ph}, \textsc{composite}}$  & $S_\text{vel}$ & SN$^a$ & SA$^b$ &  \text{F$_\text{10.7-cm}$$^c$} & Ca \textsc{ii} K$^d$ & $\Delta\nu$$^e$ \\    
\hline                  
SN$^{a}$ &  0.86 & 0.84 & 0.86 & 0.84 & 0.88 & n/a&  0.98 & 0.99 & 0.98 & 0.92\\
SA$^b$ & 0.91 & 0.89 & 0.90 & 0.89 & 0.90 & $-$ & n/a & 0.98 & 0.96 & 0.93\\
\text{F$_\text{10.7-cm}$}$^{c}$  &  0.88 & 0.86 & 0.88 & 0.86 & 0.90 & $-$ & $-$ & n/a & 0.98 & 0.94\\
Ca \textsc{ii}\,K$^{d}$  &  0.87 & 0.85 & 0.86 & 0.85 & 0.87 & $-$ & $-$ & $-$ & n/a & 0.93 \\
$\Delta\nu^e$ & 0.89 & 0.87 & 0.89 & 0.88 & 0.86 & $-$ & $-$ & $-$ & $-$ & n/a\\
 \hline                                
\end{tabular}
\tablefoot{$^a$Total sunspot number; $^b$ Total sunspot area; $^c$ 10.7-cm radio flux; $^d$ Ca  \textsc{ii}\,K-line emission index; and $^e$ Mean $l=0,1$, and 2 frequency shifts}
\end{minipage}
\end{table*}
%----------------

%----------------
\begin{table*}[ht]
\begin{minipage}{\textwidth}
\caption{Spearman's correlation coefficients between the photometric VIRGO/SPM $S_{\text{ph}}$ magnetic proxies and standard indices of solar activity during Cycle 23.}
\label{table:table2}      
\centering               
\renewcommand{\footnoterule}{}               
\begin{tabular}{c c c c c c c c c c}        
\hline\hline   
 Activity proxy &  $S_{\text{ph}, \textsc{blue}}$ & $S_{\text{ph}, \textsc{green}}$ & $S_{\text{ph}, \textsc{red}}$ & $S_{\text{ph}, \textsc{composite}}$  & SN$^a$ & SA$^b$ &  \text{F$_\text{10.7-cm}$}$^c$ & Ca \textsc{ii} K$^d$  & $\Delta\nu$$^e$\\    
\hline

SN$^{a}$ & 0.85 & 0.83 & 0.86 & 0.84 &   n/a & 0.98 & 0.99 & 0.98 & 0.93\\ 
SA$^b$ & 0.89 & 0.88 & 0.90 & 0.88 &  $-$ & n/a & 0.98 & 0.96 & 0.93\\
\text{F$_\text{10.7-cm}$}$^{c}$  & 0.86 & 0.84 & 0.87 &  0.84 &   $-$ & $-$ & n/a & 0.98 & 0.94\\
Ca \textsc{ii}\,K$^{d}$  &  0.85 & 0.83 & 0.86 & 0.84  &  $-$ & $-$ & $-$ & n/a & 0.93\\
$\Delta\nu^e$ &  0.88 & 0.86 & 0.88 & 0.87 & $-$ & $-$ & $-$ & $-$ & n/a\\
 \hline         
\end{tabular}
\end{minipage}
\tablefoot{$^a$Total sunspot number; $^b$ Total sunspot area; $^c$ 10.7-cm radio flux; $^d$ Ca  \textsc{ii}\,K-line emission index; and $^e$ Mean $l=0,1$, and 2 frequency shifts}
\end{table*}
%----------------

%----------------
\begin{table*}[ht]
\begin{minipage}{\textwidth}
\caption{Spearman's correlation coefficients between the photometric VIRGO/SPM $S_{\text{ph}}$ magnetic proxies and standard indices of solar activity during Cycle 24.}
\label{table:table3}      
\centering               
\renewcommand{\footnoterule}{}          
\begin{tabular}{c c c c c c c c c c}        
\hline\hline   
 Activity proxy &  $S_{\text{ph}, \textsc{blue}}$ & $S_{\text{ph}, \textsc{green}}$ & $S_{\text{ph}, \textsc{red}}$ & $S_{\text{ph}, \textsc{composite}}$  & SN$^a$ & SA$^b$ &  \text{F$_\text{10.7-cm}$}$^c$ & Ca \textsc{ii} K$^d$  & $\Delta\nu$$^e$ \\    \hline
 
SN$^{a}$ & 0.92 & 0.88 & 0.86 & 0.87 & n/a & 0.98 & 0.97 & 0.93 & 0.86\\ 
SA$^b$ & 0.94 & 0.90 & 0.89 & 0.90 & $-$ & n/a & 0.97 & 0.93 & 0.86\\
\text{F$_\text{10.7-cm}$}$^{c}$  & 0.94 & 0.92 & 0.90 & 0.92 & $-$ & $-$ & n/a & 0.96 & 0.91 \\
Ca \textsc{ii}\,K$^{d}$  &   0.89 & 0.89 & 0.83 & 0.87 & $-$ & $-$ & $-$ & n/a & 0.94\\
$\Delta\nu^e$ & 0.87 & 0.88 & 0.87 & 0.89 & $-$ & $-$ & $-$ & $-$ & n/a \\
 \hline      
\end{tabular}
\end{minipage}
\tablefoot{$^a$Total sunspot number; $^b$ Total sunspot area; $^c$ 10.7-cm radio flux; $^d$ Ca  \textsc{ii}\,K-line emission index; and $^e$ Mean $l=0,1$, and 2 frequency shifts}
\end{table*}
%----------------

%________________________________________________________________
\section{Photometric and velocity activity proxies}
\label{sec:activity}
The photospheric activity proxy $S_\text{ph}$ was calculated for the VIRGO/SPM \textsc{blue}, \textsc{green}, and \textsc{red} channels independently, as well as for the \textsc{composite} data, over sub series of 125 days, corresponding to  $5\times P_{\text{rot}_\sun}$, for a solar rotation at the equator  $P_{\text{rot}_\sun}$ of 25~days. An overlap of 15.625~days between consecutive sub series was also introduced. Only non-zero data points (i.e. no missing data) were used in the computation of $S_\text{ph}$. The associated standard errors of the mean values were taken as estimates of the errors on $S_\text{ph}$. A similar analysis of the radial velocity GOLF observations provided the so-called $S_\text{vel}$ activity proxy. 

%________________________________________________________________
\subsection{Photon noise correction}
The $S_\text{ph}$ and $S_\text{vel}$ activity proxies were corrected from the associated photon noise estimated from the high-frequency region (over 8000~$\mu$Hz) of the power spectrum of the corresponding observations. Figure~\ref{fig:fig2} shows the photon noise as a function of time for VIRGO/SPM and GOLF. It was estimated from the unfiltered observations with their original temporal cadence, i.e. 60 and 10 seconds respectively (see Section~\ref{sec:obs}), in order to have access to the frequency regime of the photon noise with the highest Nyquist frequency possible. In the case of VIRGO, the cadence of 60 seconds gives a Nyquist frequency of $8333.33\,\mu$Hz, which is close to the pseudo-mode regime between the acoustic cut-off frequency at $\sim$~5500\,$\mu$Hz and about $7500\,\mu$Hz \citep{garcia98,jimenez05}. 
The power spectrum reaches the Gaussian distributed white noise regime above $8000\,\mu$Hz which allows to estimate the photon noise up to $8200\,\mu$Hz. We checked that over that frequency range the mean and the standard deviation of the power spectrum return similar values. We note that the signature of the Data Acquisition System (DAS) of the VIRGO instrument with a 3-min cadence at $5555.55\,\mu$Hz has two important harmonics at $2777.78\,\mu$Hz in the p-mode region and at the $8333.33\,\mu$Hz Nyquist frequency. For the latter reason, we thus limited the calculations of the VIRGO/SPM photon noise to the upper limit of $8200\,\mu$Hz.

For validation purposes, we also have calculated the VIRGO/SPM photon noise by performing a Bayesian analysis of the background using the DIAMONDS code \citep{corsaro14}. The background profile, $P$, of the unsmoothed power spectra was modeled as described in \citet{corsaro15} following:
 \begin{equation}
 P(\nu) = B(\nu) + L(\nu) + N,
 \label{eq:eq0}
\end{equation}
where, $B(\nu)$ includes the contribution of the different time scales of granulation and the signature of activity at low frequency, and $L(\nu)$ corresponds to the Lorentzian-like power excess envelope of the solar oscillations over the entire p-mode region. The variable $N$ represents the flat photon noise at high frequency. We note that the background of asteroseisimic targets is obtained in a similar manner \citep[see e.g.,][]{kallinger14}.  We compared $N$ to the estimates of the  photon noise obtained as the mean value of the power spectra between $8000\,\mu$Hz and $8200\,\mu$Hz as explained above. We found that the values of the photon noise thus measured correspond to within 1.0, 0.7, and 0.4\,ppm to the ones inferred from the bayesian analysis respectively for the \textsc{blue}, \textsc{green}, and \textsc{red} channels. Although they should be considered as upper limits of the VIRGO/SPM photon noise, the differences are well within the mean errors on the $S_\text{ph}$ proxy of 5, 4, and 2\,ppm respectively. Furthermore, they represent an easy and reproducible way to obtain the VIRGO/SPM photon noise.

The top panel of Fig.~\ref{fig:fig2} shows that the temporal evolution of the VIRGO/SPM photon noise is different between the three SPM channels. The \textsc{blue} channel is the noisier and shows a clear increase with time starting from the beginning of the mission for a total of about 30\% between 1996 and 2016, while over the same period the photon noise of the \textsc{green} channel increases by about 10\%. On the other hand, the photon noise of the \textsc{red} channel remains quite constant with an overall variation of less than 1\%. We interpret these results as a consequence of the degradation of the filters due to the incoming radiation that increases the photon noise level as a function of the wavelength associated to a probable degradation of the detectors. This was verified by looking at the original signals which show similar behaviors between channels. We also note a sharp increase of the photon noise in the three channels from $\sim$~2017 which could indicate a new degradation of the instrument. Future observations will help to understand its origin.  

In the case of the GOLF observations, the original temporal sampling of 10 seconds (i.e., a Nyquist frequency at $50000.00\,\mu$Hz) is short enough to have undoubtedly access to the photon noise regime. Due to its single-wing configuration, the GOLF velocity data are sensitive to the photon counting, and thus to the orbital period as we can see on the bottom panel of Fig.~\ref{fig:fig2}  with a clear 365-day signature. This dependence is added to the aging of the instrument \citep{garcia05}. The configuration changes between blue-wing and red-wing observations are also visible as denoted by the vertical dashed lines.  

For each VIRGO/SPM and GOLF datasets, the photon noise was thus determined over the same 125-day sub series and subtracted to the associated $S_\text{ph}$ and $S_\text{vel}$ activity proxies.

%________________________________________________________________
\subsection{Long-term variability}
\label{sec:11year}
The temporal evolution of the $S_\text{ph, \textsc{green}}$ and $S_\text{vel}$ activity proxies calculated over 21 years of the photometric VIRGO/SPM \textsc{green} channel and radial velocity GOLF observations respectively and covering the solar cycles~23 and 24 are shown on Fig.~\ref{fig:fig3}. We see that they follow a clear 11-year modulation being larger at times of maxima. We note also that comparable temporal variations are obtained for the $S_\text{ph, \textsc{blue}}$  and $S_\text{ph, \textsc{red}}$ proxies from the \textsc{blue} and \textsc{red} observations, but different sensitivity is observed between the three VIRGO/SPM channels. The wavelength dependence of the $S_\text{ph}$ proxy is discussed in Section~\ref{sec:wave}. We note as well that smaller values of $S_\text{vel}$ are obtained when GOLF was observing in the red wing of the sodium line in comparison of the blue wing. This is confirming that the red-wing configuration has contributions from regions higher in the solar atmosphere than the blue-wing configuration.

In the following, both the $S_\text{ph}$ and $S_\text{vel}$ proxies are compared to standard solar activity indices, whose daily values were averaged over the same 125-day sub series: (1) the total sunspot number\footnote{Source:~WDC-SILSO, Royal Observatory of Belgium, Brussels at \url{http://www.sidc.be/silso/datafiles}}; (2) the total sunspot area\footnote{Source:~\url{http://solarcyclescience.com/activeregions.html}}; (3) the 10.7-cm radio flux\footnote{Source:~National Geophysical Data Center at \url{http://www.ngdc.noaa.gov/stp/solar/solardataservices.html}}; and (4) the chromospheric  Ca \textsc{ii}\,K-line emission\footnote{Source: composite data from NSO/Sacramento Peak and SOLIS/ISS observations \citep{bertello16}}. (5) The photometric and velocity proxies are also compared to the temporal variations of the $l=0,1$, and 2 acoustic oscillation frequencies averaged over the range 2650 and 3450~$\mu$Hz  and extracted from the GOLF observations over the same 125-day sub series following the methodology described in \citet{salabert15}. We note however that the sub series around the temporary losses of SoHO in summer 1998 and beginning of 1999 with a filling factor lower than 50\% were disregarded as the p-mode frequencies cannot be estimated with sufficient accuracy. These standard proxies are sensitive to different layers of the Sun. The sunspot number and area are associated to the photospheric magnetic flux, while the radio flux is an activity proxy of the upper chromosphere and lower corona and the Ca \textsc{ii}\,K-line emission a proxy of the strength of plasma emission in the chromosphere. As for the frequency shifts, they reveal inferences on the sub-surface changes with solar activity not detectable at the surface by other activity proxies. We note as well that Ca \textsc{ii}\,K observations up to June 10, 2015 only were used here as instrumental problems need to be addressed with the latest data. These proxies are compared to $S_\text{ph}$ and $S_\text{vel}$ on Fig.~\ref{fig:fig3}.

Table~\ref{table:table1} gives the Spearman's correlation coefficients between the  photometric $S_\text{ph}$ of the three VIRGO/SPM channels, the associated {\it Kepler}-like composite VIRGO/SPM time series, and the GOLF radial velocity $S_\text{vel}$ activity proxies and the standard proxies monitoring solar activity represented in Fig.~\ref{fig:fig3}. The correlations between the common activity indices  are given as well for comparison.
In the case of GOLF, the correlation coefficients were calculated only for the blue-wing period starting on 2002 November 19 in order to avoid the change between the blue and red wings of the sodium line between 1998 and 2002. We note also that independent points only were used.
Both the $S_\text{ph}$ and $S_\text{vel}$ proxies are well correlated with the standard solar activity proxies with degrees of correlation around 0.9 with no noticeable difference between the different analyzed datasets.

% -------------------
\begin{figure*}[tbp]
\begin{center} 
\includegraphics[width=0.48\textwidth,angle=0]{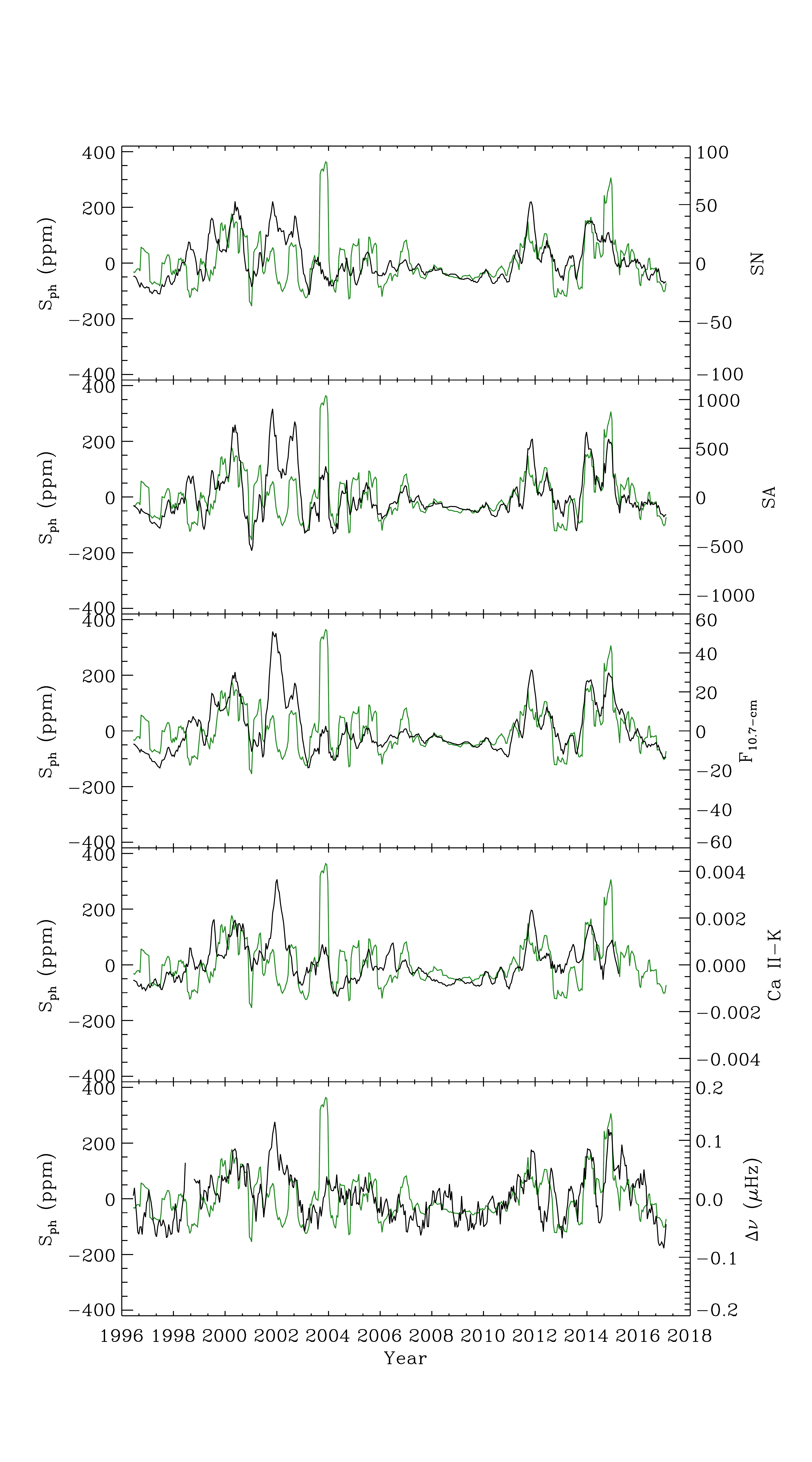}
\includegraphics[width=0.48\textwidth,angle=0]{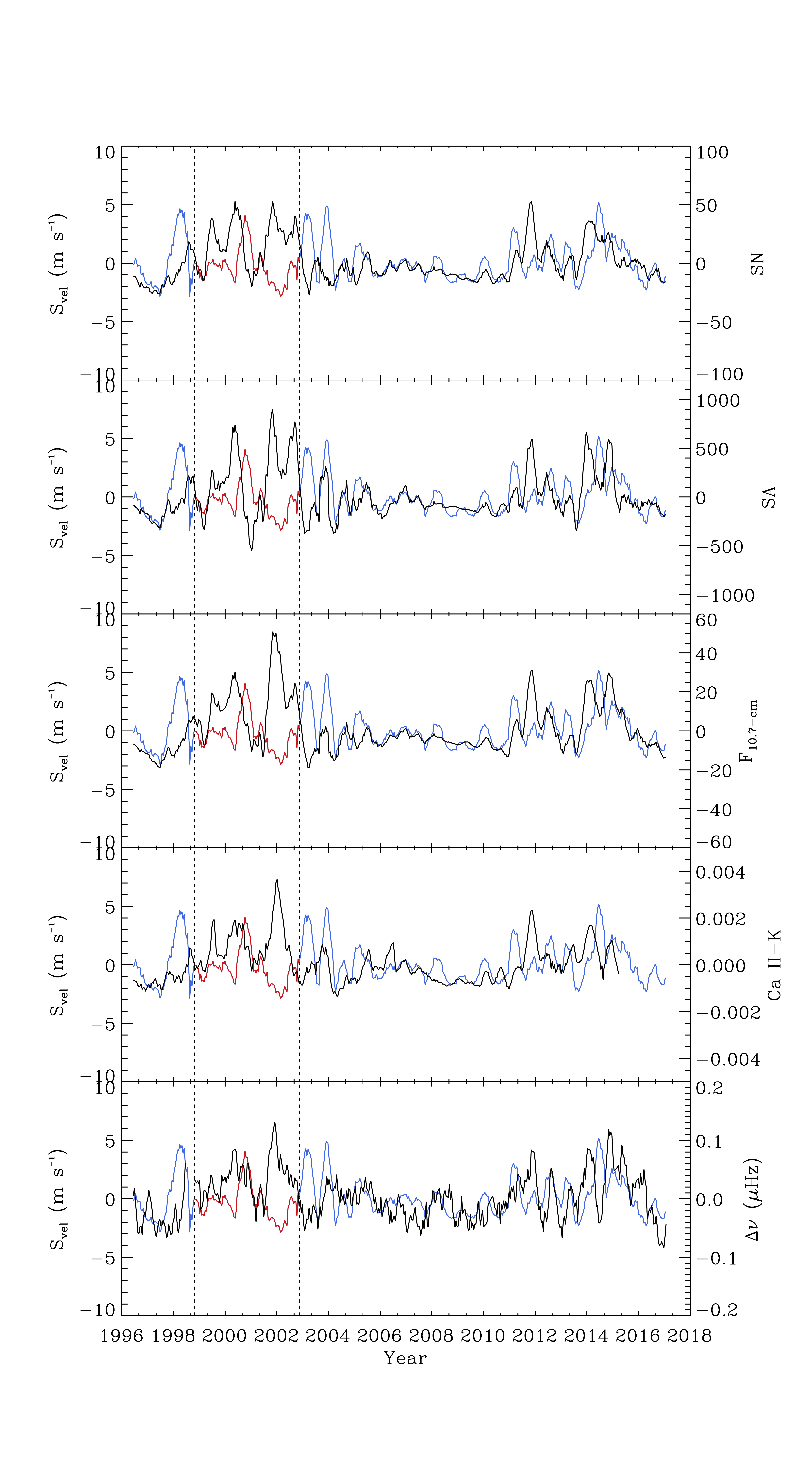}
\end{center}
\caption{\label{fig:fig4} 
Same as Fig.~\ref{fig:fig3} but once the signature of the 11-year solar cycle is removed with a Gaussian filtering (see Section~\ref{sec:qbo}). On the right panels, the two vertical dashed lines indicate the GOLF red-wing period from mid 1998 to end 2002.}
\end{figure*} 
% -------------------

Figure~\ref{fig:fig5} shows the standard activity proxies illustrated in Fig.~\ref{fig:fig3} as a function of the corresponding photospheric activity index $S_\text{ph}$ calculated from the VIRGO/SPM \textsc{green} channel observations. Similar results are obtained with the two other VIRGO/SPM channels. We note that such representation for the GOLF $S_\text{vel}$ is more difficult to interpret because the changes in the side of the observing wings. For illustrative purpose and in order to remove short-term variability, the data points were smoothed over a period of 2.5 years. The different colors correspond to different phases in solar activity: in red and blue, the rising and declining phases of Cycle~23 respectively; in green and yellow, the rising and declining phases of Cycle~24 respectively. The individual data points before smoothing are represented by the gray dots. There is a clear relationship between the photospheric index $S_\text{ph}$ and the others activity proxies from one cycle to the next. The $S_\text{ph}$ also shows a smaller variation during the weaker Cycle~24 as it observed in the other activity proxies, which is interestingly not as noticeable for the frequency shifts which are sensitive to the sub-layers of the Sun. Such differences of behavior of the frequency shifts at the beginning of the new Cycle~24 were already reported in \citet{salabert09} and \citet{tripathy10}. \citet{salabert15} also show that these differences are a function of frequency. Indeed, the frequency shifts of the low-frequency modes, less sensitive to the upper layers of the Sun, show almost no difference between Cycles 23 and 24 while the frequency shifts of the high-frequency modes are smaller during Cycle~24.
Furthermore, Fig.~\ref{fig:fig5} shows that the relationship between the $S_\text{ph}$  proxy and the standard activity proxies is not fully linear and that it follows an hysteresis pattern. Such hysteresis have already been observed between several solar observations of photospheric and chromospheric activity \citep[see e.g.,][]{bach94,ozguc12}, as well as with frequency shifts \citep[see e.g.,][]{chano98,tripathy01}. Recently, \citet{salabert16} have already shown the existence of such hysteresis between the $S_\text{ph}$ and the Ca K-line emission. Time delays are thus clearly present among the different solar activity proxies associated to the distribution in latitude of the surface magnetic flux and of its temporal evolution.

Finally, we checked if any differences can be measured between Cycle~23 and the weaker Cycle~24. Again, we focused here on $S_\text{ph}$ from VIRGO observations as the results with the GOLF $S_\text{vel}$ between Cycles 23 and 24 are more difficult to interpret because the change in the operation of the instrument throughout the mission. The obtained correlation coefficients are given in Tables~\ref{table:table2} and \ref{table:table3}. As in Table~\ref{table:table1}, independent points only were used in the calculations. The $S_\text{ph}$ proxies for the \textsc{blue} and \textsc{green} channels are observed to be better correlated with the sunspot number, the sunspot area, the radio flux, and the Ca\,\textsc{ii}\,K emission during Cycle~24 than during Cycle~23. On the other hand, the $S_\text{ph}$ proxy for the \textsc{red} channel does not show differences between the two solar cycles. That could indicate changes in the photospheric response of these activity proxies during Cycle~24 and not in the chromospheric one as the sensitivity of the \textsc{blue} and \textsc{green} VIRGO channels peaks below the photosphere.
Besides, correlations between standard activity proxies are comparable during the two last solar cycles, except with the p-mode frequency shifts and somehow the Ca\,\textsc{ii}\,K emission. Indeed, the frequency shifts show lower degree of correlation with the sunspot number, the sunspot area, the radio flux, and the Ca \textsc{ii}\,K during Cycle~24 than during Cycle~23 indicating that the sub-surface layers have not changed \citep[see][]{salabert15}. To summarize, Tables~\ref{table:table2} and \ref{table:table3} indicate that modifications of the magnetic field interaction with the solar photosphere and chromosphere, but also with the inner upper layers of the Sun, are taking place between the two last solar cycles.

%________________________________________________________________
\subsection{Short-term variability}
\label{sec:qbo}
Short-term variations in the solar activity of about 2 years (QBO) coexist on top of the 11-year cycle and are observed by intermittence with time-dependent amplitudes \citep[see e.g.,][and references therein]{bazi15}. Furthermore, these quasi-periodic variations are observed to be stronger during solar maxima. Figure~\ref{fig:fig4} shows the activity proxies $S_\text{ph}$ ($\textsc{green}$ channel) and $S_\text{vel}$, and the standard activity proxies once the signature of the 11-year solar cycle is removed. 
It was done by  substracting the corresponding smoothed variations using a Gaussian filter of full width of 2.5-year length.
 Short-term variability exists, being more or less in phase between each others. We observe also stronger short-term variations in the $S_\text{ph}$ and $S_\text{vel}$ at time of maxima as observed in other proxies. We note as well that the standard proxies are more in phase with $S_\text{ph}$ during Cycle~24 than during Cycle~23, as it was translated into their degrees of correlation given in Tables~\ref{table:table2} and \ref{table:table3}. Regarding the short-term variability of the p-mode frequency shifts, we observe a clear change in behavior of the QBO from May 2011. Indeed, a shorter variability of about 270\,days is in place for few years during the rising phase of Cycle~24 up to about February 2015 and it disappears during the following declining phase of activity. The solar cycle~24 is clearly peculiar compared to previous cycles and further in-depth analysis of the frequency shifts will be required in order to relate this to any modification of the solar dynamo.

% -------------------
\begin{figure*}[tbp]
\begin{center} 
\includegraphics[width=1\textwidth,angle=0]{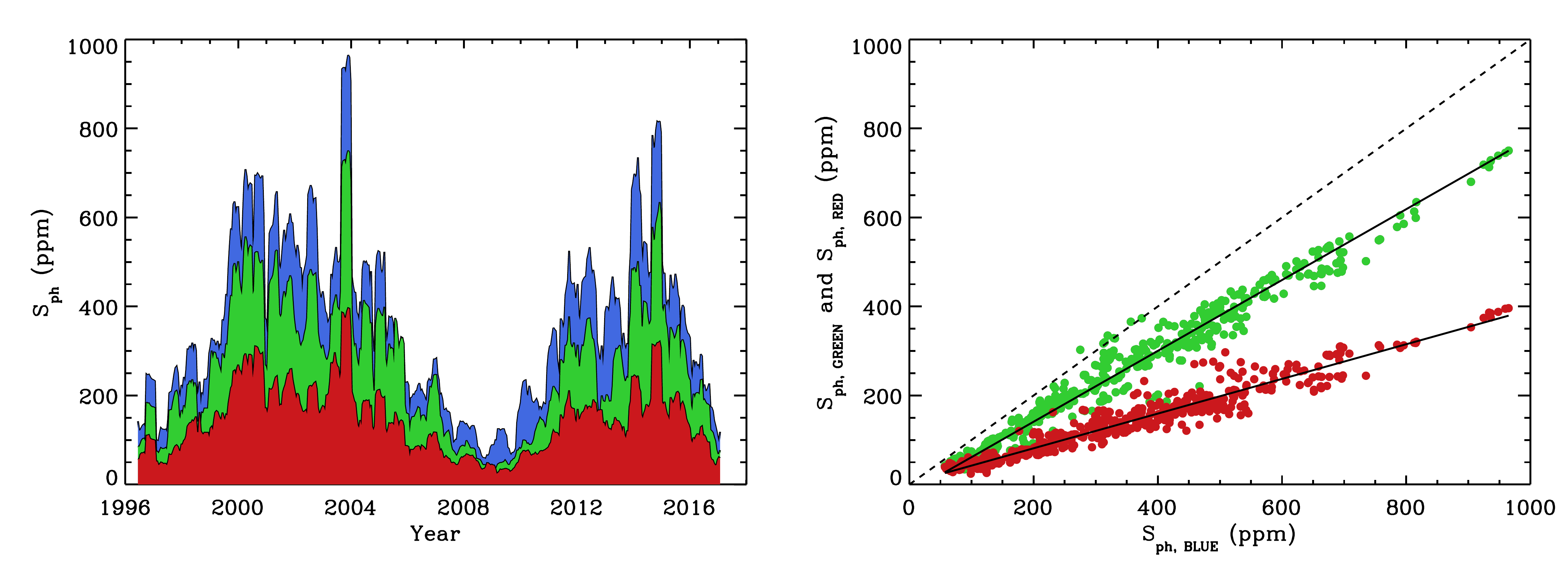}
\end{center}
\caption{\label{fig:fig10}  
{\it Left panel}: Photospheric activity proxies, $S_{\text{ph}}$ (in ppm), for each of the \textsc{blue}, \textsc{green}, and \textsc{red} channels of the VIRGO/SPM observations as a function of time. {\it Right panel}: Photospheric activity proxy, $S_{\text{ph}}$ from the VIRGO/SPM \textsc{green} (green dots) and \textsc{red} (red dots) channels as a function of the photospheric index, $S_{\text{ph}}$, from the \textsc{blue} channel. The solid lines correspond to weighted linear regressions between the data, and the dashed line represents the 1:1 correlation. }
\end{figure*} 
% -------------------

%________________________________________________________________
\section{Photometric wavelength dependence to $S_\text{ph}$ }
\label{sec:wave}
The left panel of Fig.~\ref{fig:fig10} shows the photospheric activity proxies $S_\text{ph}$ obtained from the three individual \textsc{blue}, \textsc{green}, and \textsc{red} VIRGO/SPM channels as a function of time. 
While the $S_\text{ph}$ is well correlated with solar activity for the three channels, the sensitivity of their variations along the solar cycle is different between channels. Indeed, the \textsc{blue} channel shows the largest variations between minimum and maximum of solar activity -- of about 600~ppm, while in the mean time  the \textsc{red} channel shows only a variation of about 200~ppm. The \textsc{green} channel is closer to the \textsc{blue} channel with a variation of about 500~ppm between minimum and maximum of solar activity. We note that the mean uncertainties on  $S_\text{ph}$ are about 5, 4, and 2~ppm for the \textsc{blue}, \textsc{green}, and \textsc{red} channels respectively. The largest differences between channels occur at times of solar maxima, while the differences are marginal during solar minima. 
However, the temporal variations of  $S_\text{ph}$ between the three channels remain linear, as illustrated on the right panel of Fig.~\ref{fig:fig10} which shows the $S_\text{ph}$ proxy from the \textsc{green} and \textsc{red} channels as a function of the $S_\text{ph}$ values from the \textsc{blue} channel. 
The $S_\text{ph, \textsc{red}}$  has a much reduced amplitude during the phase of maximum of activity, while comparable values are measured between the three channels during the minimum of activity. Regarding the $S_\text{ph, \textsc{green}}$, it has a much  closer temporal  behavior to the $S_\text{ph, \textsc{blue}}$ channel and to the 1:1 correlation. Weigthed linear regressions between the quantities give the following relations between the VIRGO/SPM channels:

\begin{equation}
S_{\text{ph}, \textsc{green}} = 0.756\times S_{\text{ph}, \textsc{blue}} -17.497
\label{eq:eq1}
\end{equation}
and
\begin{equation}
S_{\text{ph}, \textsc{red}} = 0.374\times S_{\text{ph}, \textsc{blue}} + 3.653
\label{eq:eq2}
\end{equation}

The Eqs.~\ref{eq:eq1} and \ref{eq:eq2} describe the wavelength dependence of the sensitivity of the $S_\text{ph}$ proxy to magnetic activity in photometric observations.
Shorter wavelengths (such as the one from the VIRGO/SPM  \textsc{blue} channel at 402~nm) are observed to be more favorable than longer wavelengths (like the VIRGO/SPM \textsc{red} channel at 862~nm) to follow solar activity. 
We note that \citet{fligge98} observed comparable relative variations in the irradiance between the three VIRGO channels. Moreover, the values given in Eqs.~\ref{eq:eq1} and \ref{eq:eq2} are in agreement with the ones derived by \citet{unruh99} using the VIRGO observations of the total and spectral irradiance. \citet{frohlich97} and \citet{jimenez99} found comparable ratios as well in the amplitude and the gain respectively between the three channels when measuring the acoustic oscillations.

These results can be related to the wavelength dependence of the magnitude of solar variability measured in sunspots. Indeed, the $S_\text{ph}$ proxy corresponds to the photometric variations associated to the rotation of spots at the surface of the Sun. Although likely to be more complex, in a first approach, we might suppose that the wavelength dependence of the $S_\text{ph}$ proxy is connected to the temperature dependence described by the Planck\textquotesingle s law explaining the larger variations observed at shorter wavelengths than at longer wavelengths. Although the studied range of wavelengths is narrow (from 402~nm to 862~nm, going from the visible light to the near infrared), the Eqs.~\ref{eq:eq1} and \ref{eq:eq2} show as a first approximation that a simple linear relation can be used to rescale the activity proxy $S_\text{ph}$ obtained at different wavelengths. It thus offers the possibility to compare stellar activity to the Sun measured from different photometric observations of solar-like and main-sequence stars. For instance, the bandpass of the {\it Kepler} telescope is comprised between 420 and 900~nm. Although wider, the bandpass of the PLATO \citep{rauer14} space mission should cover a comparable wavelength range as {\it Kepler}.
%________________________________________________________________

%________________________________________________________________
\section{Conclusions}
\label{sec:conclusion}
We analyzed the observations collected by the space-based photometric VIRGO and radial velocity GOLF instruments on board the SoHO satellite in an unexplored way yet to study the long- and short-term variabilities of solar activity. We tracked the temporal modulations of the observed parameters associated to the presence of spots or magnetic features rotating on the surface of the Sun in a similar manner as it is done in the stellar context with CoRoT and {\it Kepler}. A total of 21 years of observations spanning the solar cycles 23 and 24 were thus divided into sub series of $5\times P_{\text{rot}_\sun} = 125$\,days, with a mean rotational period of the Sun $P_{\text{rot}_\sun}$ of 25\,days. The associated photospheric activity proxy $S_\text{ph}$ was then calculated for the photometric VIRGO/SPM \textsc{blue}, \textsc{green}, and \textsc{red} channels independently, as well as for the combination of the \textsc{green} and \textsc{red} channels into a \textsc{composite} time series. This \textsc{composite} dataset has the closest bandwidth to {\it Kepler} observations and should be used when comparing the Sun to {\it Kepler} targets. The corresponding activity proxy $S_\text{vel}$ was also determined from the radial velocity GOLF observations. 

Both $S_\text{ph}$ and $S_\text{vel}$ were compared to several standard solar activity proxies which are sensitive to different layers of the Sun, from the inner sub-layers up to the corona: the temporal variability of the acoustic oscillation frequencies, the sunspot number and sunspot area, the Ca \textsc{ii}\,K-line emission, and the 10.7-cm radio flux. We showed that these new activity proxies derived from the VIRGO and GOLF observations are well correlated with commonly used activity proxies over the long-term 11-year variability and that they indicate as well a weaker Cycle~24 compared to the previous solar cycle as the other proxies do. Moreover, these new proxies suggest that modifications of the magnetic field interaction with the solar photosphere and chromosphere, but also with the inner upper layers of the Sun,  took place between Cycle~24 and Cycle~23. The short-term variability observed in the Sun, such the QBO,  was also measured to be comparable to other proxies with stronger variations at time of maxima. We demonstrated that these new proxies provide a new manner to monitor solar activity in photometry and in radial velocity observations.

Furthermore, the photometric proxies $S_\text{ph}$ show a wavelength dependence  between the three channels of the VIRGO photometers. The \textsc{blue} channel at 402~nm is the most sensitive to variations with solar activity, while the \textsc{red} channel at 862~nm is the least sensitive by a factor 2.7. The \textsc{green} channel at 500~nm has closer activity sensitivity to the \textsc{blue} channel with a factor 1.3. Such observations of the wavelength dependence of the response function of the solar photosphere provide inputs for the study of the stellar magnetism of Sun-like stars. It allows also to rescale the activity proxy $S_\text{ph}$ of solar-like and main-sequence stars obtained at different wavelengths with the photometric {\it Kepler} and PLATO observations and to compare to the Sun. These measurements of solar magnetic activity monitoring derived from the VIRGO and GOLF observations are deliverables of the European SpaceInn project, and can be found at {\url{http://www.spaceinn.eu}}. 
%________________________________________________________________

%________________________________________________________________
\begin{acknowledgements}
The GOLF and VIRGO instruments on board SoHO are a cooperative effort of many individuals, to whom we are indebted. SoHO is a project of international collaboration between ESA and NASA. The authors strongly acknowledge the french space agency, CNES, for its support to GOLF since the launch of  SoHO. We are also particularly grateful to Catherine Renaud for her daily check of the GOLF data. DS acknowledges the financial support from the CNES GOLF grant and the Observatoire de la C\^ote d'Azur for support during his stays. EC is funded by the European Union$^\prime$s Horizon 2020 research and innovation programme under the Marie Sklodowska-Curie grant agreement no. 664931. The research leading to these results has also received funding from the European Community's Seventh Framework Programme ([FP7/2007-2013]) under grant agreement no. 312844 (SPACEINN). The sunspot number data are provided by WDC-SILSO, Royal Observatory of Belgium (Brussels). We thank Solar Cycle Science (\url{http://solarcyclescience.com}) for making their data of the sunspot area freely available. The 10.7-cm radio flux measurements are provided by the National Geophysical Data Center. This work uses SOLIS/ISS data obtained by the NSO Integrated Synoptic Program (NISP), managed by the National Solar Observatory, which is operated by the Association of Universities for Research in Astronomy (AURA), Inc. under a cooperative agreement with the National Science Foundation. 
\end{acknowledgements}	
%________________________________________________________________

%-------------------------------------------------------------------

\end{document}